 \documentclass[preprint,3p,times,twocolumn]{elsarticle}

\usepackage{amssymb}

\journal{Physics Letters B}

\begin{document}

\begin{frontmatter}



\title{Reanalysis of the {\sc Gallex} solar neutrino flux and source
experiments$^\dagger$}


\author{F. Kaether\corref{cor1}}
\cortext[cor1]{Corresponding author. Tel.: +49 6221 516828}
\ead{Florian.Kaether@mpi-hd.mpg.de}
\author{W. Hampel}
\author{G. Heusser}
\author{J. Kiko}
\author{T. Kirsten}
\address{Max Planck Institute for Nuclear Physics, P.O. Box 103980,
D-69029 Heidelberg, Germany}

\begin{abstract}
After the completion of the gallium solar neutrino experiments at the Laboratori
Nazionali del Gran Sasso ({\sc Gallex}: 1991-1997; GNO: 1998-2003) we have retrospectively
updated the {\sc Gallex} results with the help of new technical data that were impossible
to acquire for principle reasons before the completion of the low rate measurement
phase (that is, before the end of the GNO solar runs). Subsequent high rate
experiments have allowed the calibration of absolute internal counter efficiencies
and of an advanced pulse shape analysis for counter background discrimination. The
updated overall result for {\sc Gallex} (only) is $73.4^{+7.1}_{-7.3}$  SNU. This is 5.3\% below
the old value of $77.5^{+7.5}_{-7.8} \,\rm SNU$ \cite{Gx499}, 
with a substantially reduced error. A similar reduction is obtained
from the reanalysis of the $^{51}$Cr neutrino source experiments of 1994/1995.
\end{abstract}


\end{frontmatter}
\renewcommand{\thefootnote}{\fnsymbol{footnote}} 
\footnotetext[2]{Dedicated to Michael Altmann, Nicola Ferrari and Keith Rowley.}


\newcommand{\figwid}{7.8cm}  

\section{Introduction}
The {\sc Gallex} detector at the Gran Sasso Underground Laboratory
(LNGS) in Italy has monitored solar neutrinos with energies above 233
keV from 1991 to 1997 by means of the inverse $\beta$-decay reaction
$^{71}$Ga($\nu_e$,$e^-$)$^{71}$Ge
\cite{Gx499}\cite{Gx192}\cite{Gx1c93}\cite{Gx295}\cite{Gx396}.
Together with the subsequent GNO experiment solar neutrinos
have been recorded from 1991 to 2003, with a break in 1997
\cite{Gno00}\cite{Gno05}. The experimental procedure for a typical
{\sc Gallex} or GNO solar neutrino run has been as follows:
30.3 t of gallium in the form of a concentrated GaCl$_3$-HCl solution
are exposed to solar neutrinos for a time period between three and
four weeks. In the solution, the neutrino-induced $^{71}$Ge atoms as
well as the inactive Ge carrier atoms added to the solution at the
beginning of a run form the volatile compound GeCl$_4$, which at the
end of an exposure is swept out of the solution by means of a
nitrogen gas stream. The nitrogen is then passed through a gas
scrubber where the GeCl$_4$ is absorbed in water. The GeCl$_4$ is
finally converted to GeH$_4$ which together with xenon is introduced
into a proportional counter to determine the number of $^{71}$Ge
atoms by observing their radioactive decay (half-life 11.43 d \cite{Ham85}).

In order to reduce the background in $^{71}$Ge counting with
proportional counters the pulses recorded by the data acquisition
system were analyzed by a pulse shape discrimination method. In
contrast to GNO, the published {\sc Gallex} data have so far
been analyzed with a rather simple procedure, where pointlike ionizations 
are distinguished from extended background events by the time in which the
proportional counter signal rises from 10\% to 70\% of the amplitude
recorded by the transient digitzer. A more sophisticated method has
been developed \cite{Urb89}\cite{Sch93}\cite{Alt96a}
and tested already in {\sc Gallex} \cite{Alt96}.
However, in order to determine the cut efficiency for such a
procedure, a calibration data set with high statistics measured with
the full counting system is required. In order not to damage the low 
counter backgrounds, such data could only be acquired at the very end 
of {\sc Gallex} in the frame of the
$^{71}$As experiment \cite{GxA98}, in which a rather large number of
$^{71}$Ge decays ($\sim 10^4$) has been recorded. Using this data
set, a new pulse shape discrimination method has now been developed and
applied to the {\sc Gallex} data \cite{Kae07}.

There are two additional motivations to re\-analyse the {\sc Gallex}
solar neutrino data as well as the data from the two $^{51}$Cr
neutrino source experiments that were performed in {\sc Gallex}
\cite{GxS195}\cite{GxS298}. At first, 10 out of 22 counters used in
the {\sc Gallex} solar neutrino measurements and 4 out of 14 counters
used in the {\sc Gallex} $^{51}$Cr neutrinos source experiments have
been absolutely calibrated in the frame of the GNO experiment
\cite{Gno05}\cite{Kae03}. Secondly, there is now an improved value for the solar
neutrino signal and its error available which has to be subtracted
from the measured signal in the analysis of the $^{51}$Cr data.

\section{Pulse shape analysis in $^{71}$Ge counting} \label{psa}

$^{71}$Ge decays back to $^{71}$Ga by K, L or M electron capture. The hole
in the corresponding shell is filled by transitions of electrons
from higher shells. The released energy is mostly transferred to
electrons from the same or higher shells which subsequently are
emitted as Auger-electrons. Only in the case of L to K transitions
a substantial fraction of cases leads to the emission of a K-alpha
X-rays (9.3 keV) because of the rather high fluorescence yield of
the K shell (0.528). The range of Auger-electrons in the counter
gas is rather small ($< 1 \, \rm mm$) and therefore the volume extension
of the energy depostion is always small. On the other hand, the
mean free path of a 9.3 keV X-ray is about 1 cm and hence similar
to the counter dimensions. The X-ray is therefore
either able to leave the counter undetected or to produce a second
separated energy deposit where the ratio of the two energies is at a
fixed value of $\approx 8$. Neglecting M events (which are below
the selected energy threshold) this leads to three different kinds of events: (a)
a single electron cloud corresponding to an energy of 10.4 keV, (b) a
single electron cloud of about 1.2 keV, and (c) two electron clouds
of 1.1 keV and 9.3 keV, respectively. Contrary, background
events are mainly caused by higher energy electrons coming from beta
decays or they are induced by gamma rays via Compton effect.
These events don't produce pointlike ionizations but an ionization track
in the counting gas which leads to a slower rise time of the signals. An identification
of pointlike ionizations, double ionizations or extended (multiple) events
therefore allows to distinguish
in many cases between $^{71}$Ge decays and background events.

The new pulse shape discrimination method described here is performed
in three steps. At first, the original pulses are slightly smoothed.
This is necessary due to electronical and digital noise affecting the
pulse shape, particularly for low energy events. A piecewise
polynomial fit was used. For each data point $P(t_i)$ a region  of
$t_i \pm 8 \rm \, ns$ (corresponding to 20 data points on each side)
was fitted with a second order polynomial $p(t)$. Finally each data
point $P(t_i)$ was replaced by $p(t_i)$. This method has the
advantage that it provides an adequate noise reduction but conserves
even sharp structures on bigger time scales.

A pointlike energy deposition in the counter leads to a cloud of
primary electrons which is $\delta$-shaped (neglecting diffusive
effects) when reaching the gas amplification zone in the proportional
counter. Under ideal conditions (perfect radial electric field,
constant ion mobility) the shape of the resulting preamplifier output
pulse can be written as $P_{\delta}(t) = V_0 \log(1+t/t_0)$
\cite{Alt96}. A general pulse shape can then be described by a
convolution of the pulse shape caused by a pointlike charge cloud
with a function $j(t)$ which parameterises the number of electrons
arriving at the gas amplification zone as a function of time: $P(t) =
P_{\delta}(t) \otimes j(t)$. In order to reveal $j(t)$ from a
measured pulse $P(t)$ one has to numerically deconvolute
$P_{\delta}(t)$ from $P(t)$. This is the second step in the applied
pulse shape analysis and is performed by a Fourier analysis (i.e.
transforming the measured pulse into the frequency domain) where
deconvolution is simply a division (for more details see
\cite{Kae07}).

\begin{figure}[t]
\includegraphics[angle=0,width=\figwid]{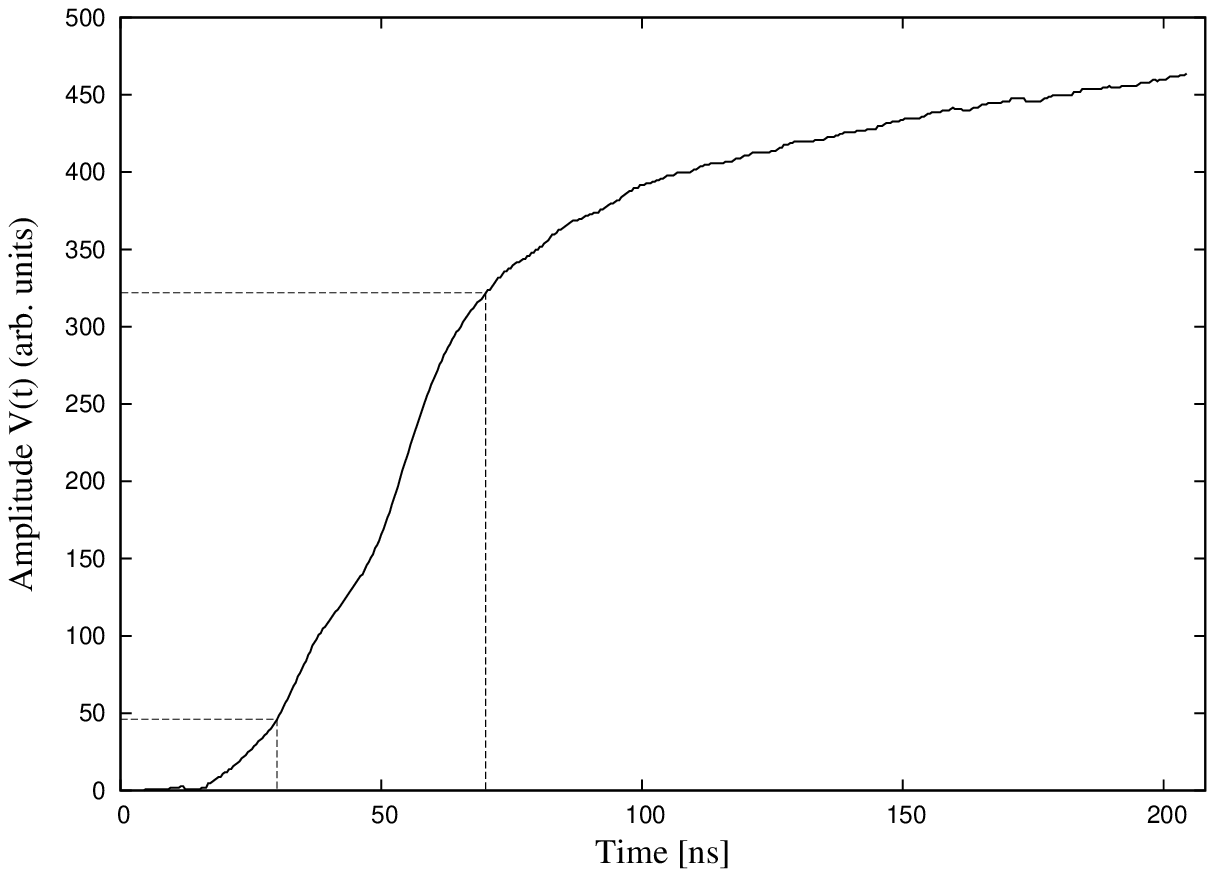}
\includegraphics[angle=0,width=\figwid]{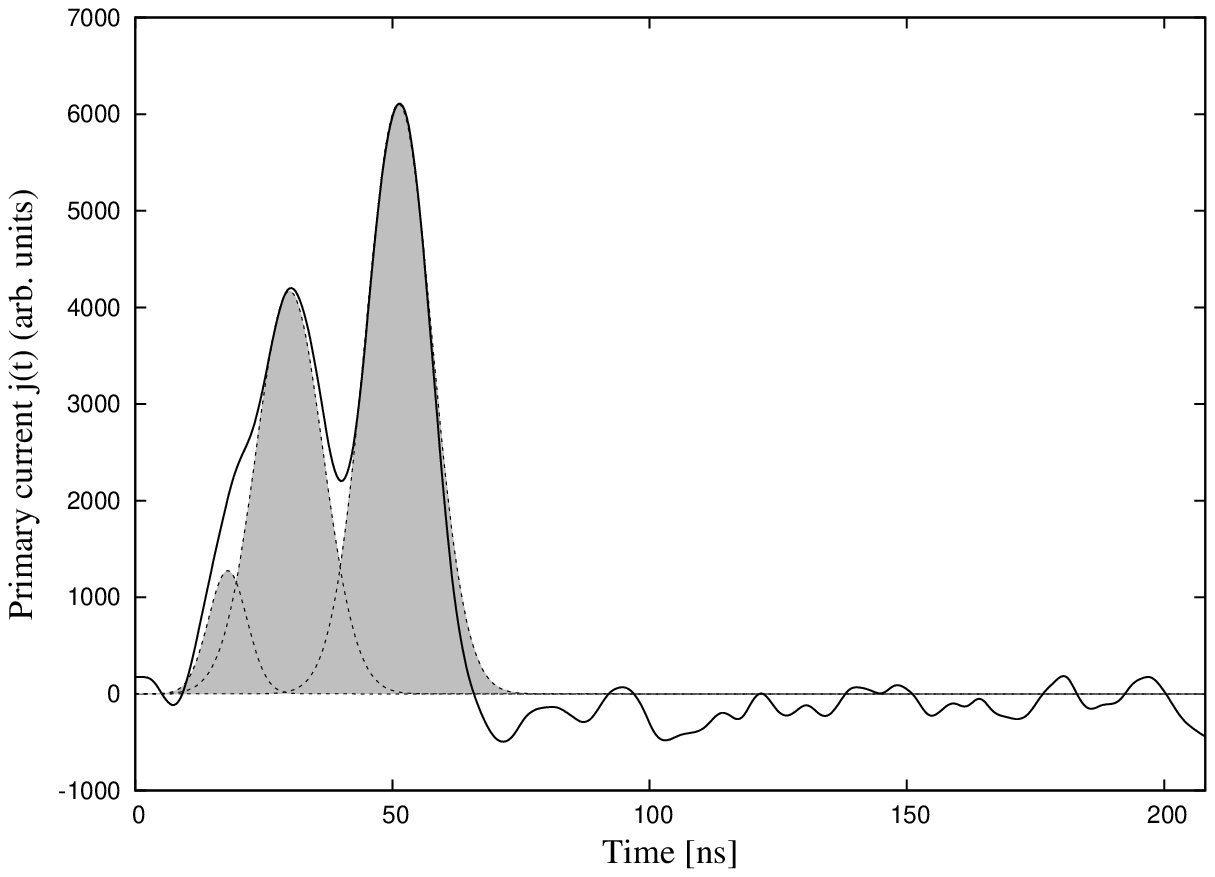}
\caption{Proportional counter signal $P(t)$ of a typical multiple
event in addition with the 10\%-70\% rise time levels (above) and the primary
current $j(t)$ of the same event derived by deconvolution with the
three major peaks (below).   \label{peaks} }
\end{figure}
\begin{figure}[t]
\includegraphics[angle=0,width=\figwid]{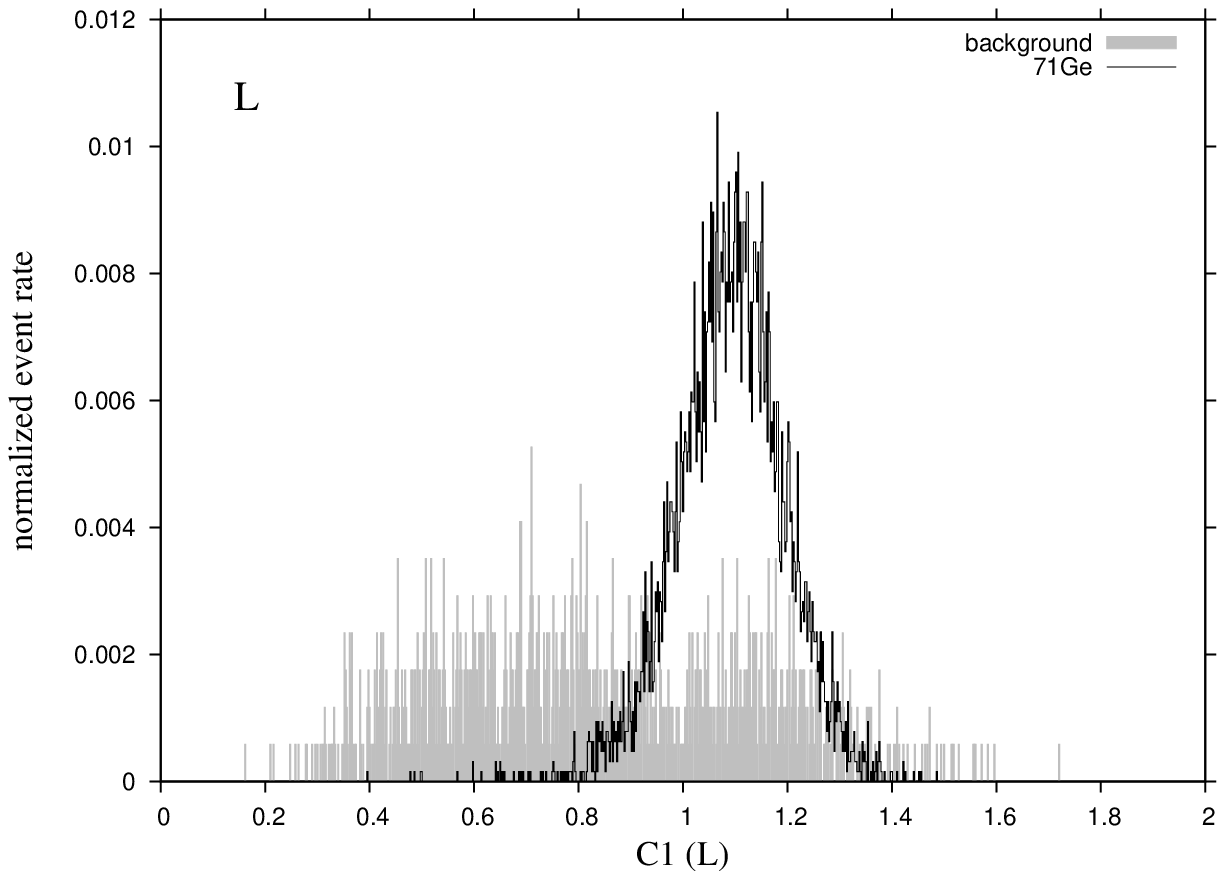}
\includegraphics[angle=0,width=\figwid]{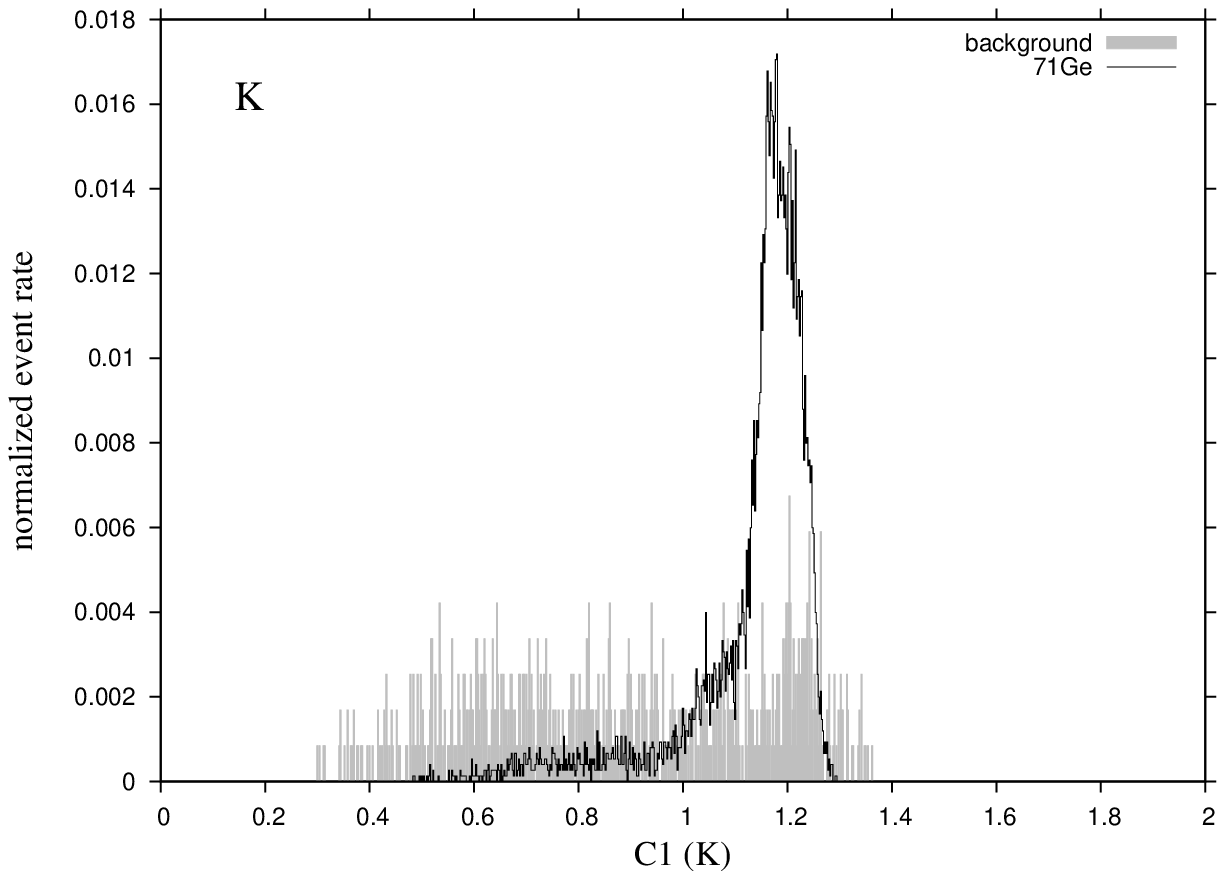}
\caption{Distribution of pulse shape parameter $C_1$ for
$^{71}\rm Ge$ events (black) and background events (grey) in the L
(above) and K (below) energy range. \label{C1-ge-bck}}
\end{figure}

An example of a typical multiple background event is shown in Figure
\ref{peaks} where the measured signal $P(t)$  and the primary current
$j(t)$ derived by deconvolution is shown in the upper and lower
figure, respectively. $j(t)$ is directly connected to the radial
charge distribution in the proportional counter and each peak is
caused by one single charge cloud. Identification of the major peaks
of $j(t)$ (see lower part of Figure \ref{peaks}) is the third step in
pulse shape analysis and is performed as follows:
\begin{list}{$\bullet$}{\setlength{\topsep}{1ex}
\setlength{\itemsep}{-0.3ex} \setlength{\leftmargin3.0ex}}
\item Determination of the maximum position $t_{\rm max}$.
\item Determination of the full width at half maximum (FWHM).
In cases of asymmetry on each side the half width was determined and
the smaller value was choosen.
\item The peak was approximated as a Gaussian $g(t_{\rm max},
\sigma)$ where ${\rm FWHM} = 2.35 \sigma$.
\item Subtraction of Gaussian and repeat the procedure.
\end{list}
The resolution of this peak search algorithm was defined as follows:
if the distance between the maxima of two peaks was smaller than the
mean of the half widths both peaks were combined into a single peak.
Regarding Figure \ref{peaks}, the distance of the two leftmost peaks is slightly
above the resolution threshold.

The total deposited energy is proportional to the number of primary
charges and therefore to the total integral $\int j(t) \, dt$. The
fraction $C_i$ of energy deposited in one single charge cloud is
therefore given by the peak integral normalised with the total energy
\begin{equation}%
 C_i = \frac{\int g(t_{\rm max}, \sigma) \, dt}{\int j(t) \, dt} \, .
\end{equation}%
For single events, where the energy deposit is concentrated in one
single charge cloud, one expects a ratio of $C_1 \approx 1$ while
$C_2$ and $C_3$ are caused by noise and therefore are small. Actually
$C_1$ is often even slightly larger then 1 due to the fact that the
negative noise part decreases $\int j(t) \, dt$. In contrast, for
multiple events, $C_1$ is obviously smaller then 1 with a
simultaneous increase of $C_2$ and $C_3$. In the case of K double
events  one expects to recover the given ratio $C_1/C_2 \approx 8$.

Following these expectations, criteria for event selection were
defined. To decide whether a parameter is suitable to distinguish
background from $^{71}\rm Ge$ events one needs reference pulses for
both kinds of events. A sample of background events can be obtained
from the solar runs themselves, since each sample was measured for
about 180 days, but after 50 days ($\approx 3\tau$) the $^{71}\rm Ge$
atoms initially present are decayed away.

A large amount of $^{71}\rm
Ge$ events were provided by the arsenic experiment \cite{GxA98}
allowing to collect the parameter distribution with good statistics.
Figure \ref{C1-ge-bck} shows the distributions of parameter $C_1$ for
$^{71}\rm Ge$ decays and background events. It is obvious that an
adequate constraint on $C_1$ allows to select $^{71}\rm Ge$ decays
and to reject a large part of background events.

Finally, a comparison with events from calibrations with an external X-ray source
(cerium) which were performed for all solar runs \cite{Gx192} provides the
individual pulse shape parameter bounds for each run and a precise
determination of the pulse shape cut efficiencies. The $C_1$
distribution for cerium events is very similar to germanium events. 
For each calibration the location and
width of the $C_1$-peak is estimated and an acceptance window for
$^{71}$Ge events is defined. The efficiency of this cut was
determined using the arsenic runs for L events to $\varepsilon_L =
0.960 \pm 0.006$. The efficiency for K events is about 80\% due to
the fact that the $C_1$ cut rejects nearly all of the double events.
To increase the number of accepted K double events an additional cut
was defined using the ratio $C_1/C_2 $. Due to the limited energy
resolution of the proportional counters a wide acceptance range for
this ratio has been choosen ($5 < C_1/C_2 < 12$). To reject multiple
background events with a ratio of $C_1/C_2$ within these bounds, an
additional upper limit for $C_3$ was defined. Altogether one obtains
an efficiency for K events of $\varepsilon=0.861 \pm 0.018$. For more
details see \cite{Kae07}.

\section{Solar run analysis}

\subsection{Event selection} \label{eventselection}

In a first step, all obvious background events are removed by
several cuts. These cuts are identical to those described in
\cite{Gx192}, except for the pulse shape cut, which was applied
according to the procedure described in the previous section.

\begin{figure}[t]
\includegraphics[angle=0,width=\figwid]{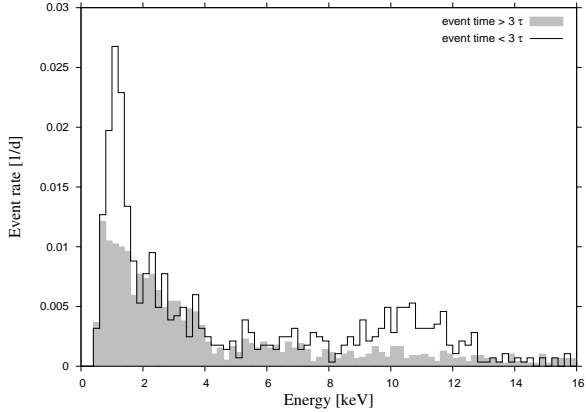}
\caption{All candidate events, divided in early
($t<3 \tau $) and late ($t>3 \tau $)
events.\label{e-spec-early-late}}
\end{figure}

All remaining candidate events (without energy cut) are plotted in
Figure \ref{e-spec-early-late}, divided into early ($t<3 \tau $) and
late events. The characteristics of a typical $^{71}\rm \, Ge$ energy
spectrum with the two peaks at 1.2 and 10.4 keV, respectively, are
quite obvious in the early spectrum (solid line). The peak positions
and widths as well as the intensities of both peaks are lying within
the expected ranges.

\subsection{Maximum Likelihood Analysis}

The final cut to the data is the energy cut, by which only events are
selected which are inside the L and K energy windows (see
\cite{Gx192}). After this cut there remain 726 and 452 events for the
L and the K energy window, respectively. These events were used for a
maximum likelihood analysis which was described  in \cite{Cle83} for
the chlorine experiment and was adapted  for  {\sc Gallex} and GNO.
The total production rate $P$ of $^{71}$Ge is
\begin{equation}
P = P_\odot / d_r^2 + P_{\rm fix} \end{equation} where $P_\odot$ is
the solar production rate which has to be corrected by the individual
Earth-Sun distance for each run $d_r$ (given in units of 1 AU), and
$P_{\rm fix}=(0.039 \pm 0.011)$ atoms per day which is a fixed
component caused by side reactions (see \cite{Gx396}). $P_\odot$ is
one of the free parameters of the likelihood function $\cal L$. In
addition one assumes the background rates in the two energy windows
$b_L$ and $b_K$ as independent free parameters for each of the 65 {\sc Gallex} runs.
Altogether the likelihood function has to be maximised for 131 free
and independent parameters. This is done by using the Fortran library
{\sc Minuit} provided by {\sc Cern} to minimise $- \log \cal L$. The
combined result for all {\sc Gallex} runs is
\begin{equation}  P_\odot = \left[ \, 73.4^{+6.1}_{-6.0} \,
{\rm (stat.)} ^{+3.7}_{-4.1} \, {\rm (syst.)}\, \right] \; {\rm SNU}
\; . \label{psolar-psa} \end{equation}%
The statistical error determination is given in maximum likelihood
theory by a variation of $\hat P_\odot$ until
\begin{equation}
 \log {\cal L}_{\rm max} - \log {\cal L}(\hat P_\odot) =  \frac{1}{2} 
\end{equation}%
while $\log {\cal L}(\hat P_\odot)$ was maximised regarding the remaining
free  parameters which leads to $1\sigma = P_\odot({\cal L}_{\rm max}) - \hat P_\odot$. A possible asymmetry of the error is considered by
investigation of both sides of ${\cal L}_{\rm max}$.

\begin{table}[t]
\center{
\renewcommand{\arraystretch}{1.2}
\begin{tabular}{ll} \hline
Efficiencies \cite{Gno05} & $\pm 2.6\%$
\\ Energy cut  \cite{Gx192} & $\pm 2.0\%$ \\ Pulse shape analysis
 & $\pm 2.0\%$ \\ Chemical yield \cite{Gx499} & $\pm
2.1\%$
\\ Target mass \cite{Gx1c93} & $\pm 0.8\%$ \\ $^{68}\rm Ge$
correction \cite{Gx499} & $^{+0.9}_{-2.6}\%$ \\ Side reactions
\cite{Gx499}\cite{Cri95}\cite {Cri97} & $\pm 1.5\%$ \\ Rn cut \cite{Gx499} & $\pm 1.5\%$ \\
\hline Sum & $ ^{+5.0}_{-5.6}\%$ \\ \hline%
\end{tabular}%
}
\caption{Systematic error contributions \label{systerr}}
\end{table}%

The systematic error includes the uncertainty of counter
efficiencies, which decreased to 2.6\% due to the more precise
calibrations \cite{Gno05}. The error of the pulse shape cut
efficiency was estimated to 2.0\%. The contribution of other
components are unchanged compared to previous publications, a
compilation is given in Table \ref{systerr}.

For the maximum likelihood analysis the half-life of $^{71}\rm Ge$ is usually
fixed to its known value of 11.43 d. However, it  can also be treated
as an additional free parameter. This yields 10.3 $\pm$ 1.2 d, which
is in agreement with the expected value. Moreover,
due to the radon cut inefficiency and the short half-life of
$^{222}\rm Rn$ and its daughters one expects a small bias
towards a shorter half-life. Besides the energy spectrum characteristics,
this is a strong proof of the {\sc Gallex} data set
consistency.

For a comparison with the previously published results we repeated the rise
time analysis. The event selection procedure described in section
\ref{eventselection} was used identically except the pulse shape
analysis was replaced by the rise time method. The new counter
efficiencies were considered as well as the correction due to the
earth-sun distance variation (which so far had not been applied in
the {\sc Gallex} data analysis). 
The result
\begin{equation}  P_\odot^{\scriptscriptstyle \rm RT}  = \left[ \, 77.4^{+6.4}_{-6.2}
\, {\rm (stat.)} ^{+3.9}_{-4.3} \, {\rm (syst.)}\, \right] \; {\rm
SNU} \;  \end{equation} is in very good agreement with the value of $\left[ \, 77.5 \pm 6.2 ({\rm stat.})^{+4.3}_{-4.7}({\rm syst.}) \, \right]$ SNU
given in \cite{Gx499}. All changes average to near zero, except for 
the pulse shape analysis. 

\begin{figure}[t]
\includegraphics[angle=0,width=\figwid]{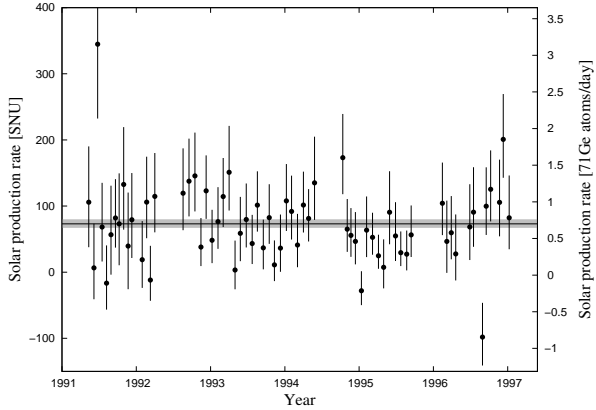}
\caption{Single results of the 65 {\sc Gallex} solar runs (error bars
are $\pm 1\sigma$ statistical).\label{davisplot}}
\end{figure}%

\subsection{Single runs and G{\footnotesize ALLEX}  I-IV}

\begin{figure}[t]
\includegraphics[angle=0,width=\figwid, height=5.6cm]{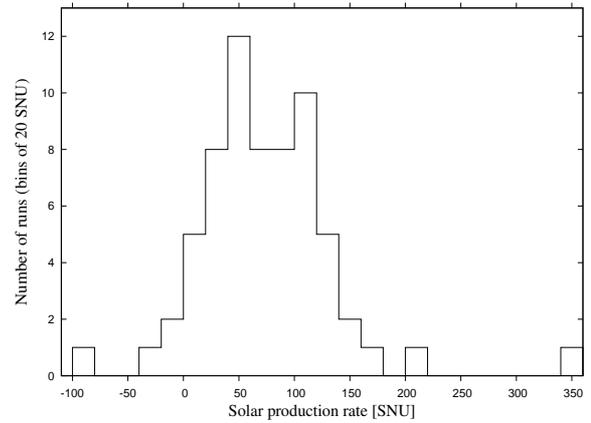}
\caption{Distribution of the {\sc Gallex} single run results in bins
of 20 SNU.\label{runhist}}
\end{figure}

The single run results are listed in Table \ref{results12} and Table \ref{results34} and are plotted 
in Figure \ref{davisplot}. The histogram 
in Figure \ref{runhist} shows the distribution of results in bins of 20 SNU.

Even though the statistical error of a single run result is usually asymmetric, one expects a normal distribution as shown by Monte Carlo simulations \cite{Gx499}. This expectation was tested by a Kolmogorow-Smirnov-test. The test value is defined as the maximum deviation $D$ between the cumulative distribution functions of the given data set and the expected normal distribution. One obtains $D=0.076$ for the {\sc Gallex} data set. For randomly generated samples one gets higher values of $D$ in 54\% of all cases and a 90\% confidence level of $D_{90}=0.1$. For a second test (which is related to the latter one but more sensitive concerning outliers) the test value was defined as the total integral of absolute deviations between the cumulative distribution functions. In 19\% of cases randomly generated samples created higher values than the original data set.  
From these points of view there is no reason to doubt the hypothesis of a normal distribution.

The statistical errors of single runs are rather big, because even a single 
accepted or rejected event is able to change the result of a run by 10 SNU or even more. 
Therefore a run by run comparison between pulse shape and rise time analysis is not very 
meaningful. Only  combinations of many runs are suitable to provide enough statistics to 
decrease the error to a significant level. Therefore  the 65 runs were sorted into groups. 
For historical reasons we stayed with the grouping in four periods of data taking which 
occured in a natural way by interruptions for
construction works or source experiments. Nevertheless, this kind of grouping is
arbitrary and should have no effect on the results.

\begin{table}[b]
\center{
\renewcommand{\arraystretch}{1.5}
\begin{tabular}{|c|c|c|} \hline
{\sc Gallex}     &  \multicolumn{2}{|c|}{Results [SNU]}     \\
period           &  rise time        & pulse shape  \\ \hline
I                &  $84.0^{+17.6}_{-16.7}$           & $75.1^{+17.3}_{-16.2}$  \\
II               &  $77.2^{+9.9}_{-9.5}$             & $82.8^{+10.0}_{-9.5}$  \\
III              &  $51.2^{+10.8}_{-10.0}$           & $49.5^{+10.7}_{-9.8}$  \\
IV               &  $122.1^{+18.4}_{-17.5}$          & $89.2^{+16.6}_{-15.5}$  \\ \hline
\end{tabular}
\caption{Results of the {\sc Gallex} periods I-IV with rise time and pulse shape analysis. The errors are 1$\sigma$ (stat.).
\label{tab-gallex1234} }
}
\end{table}%

\begin{figure}[t]
\includegraphics[angle=0,width=\figwid]{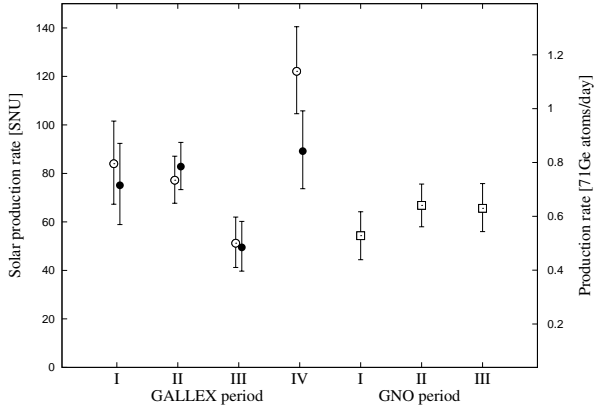}
\caption{Results of the {\sc Gallex} periods (rise time ($\scriptstyle \odot$) and pulse shape analysis ($\bullet$)) compared to the three GNO periods ($\scriptstyle \boxdot$). \label{gallex+gno}}
\end{figure}

The results of the four {\sc Gallex} periods are shown in Figure \ref{gallex+gno} 
and are listed in Table \ref{tab-gallex1234}
with rise time and pulse \hfill shape \hfill analysis, \hfill respectively. \hfill While \hfill the \hfill results \hfill of
\onecolumn

\begin{table}[h]
\begin{center}
\renewcommand{\arraystretch}{1.3}
 \footnotesize
\begin{tabular}{|cc|cc|ccc|} \hline
\multicolumn{2}{|c}{\sc Gallex I} &  \multicolumn{2}{|c|}{Exposure}  &  \multicolumn{3}{c|}{pulse shape analysis} \\ 
\multicolumn{2}{|c|}{Runs}         &  start  & duration (d)          &  $b_L$  & $b_K$ & $P_\odot$ (SNU)  \\ \hline 
1 &    b29 &  14-MAY-1991 & 21.0 &   0.028 & 0.000 & $ 105^{+  84}_{ -68}$ \\ 
2 &    b31 &   5-JUN-1991 & 20.8 &   0.020 & 0.034 & $   6^{+  67}_{ -47}$ \\ 
3 &    b32 &  26-JUN-1991 & 21.0 &   0.115 & 0.057 & $ 344^{+ 128}_{-112}$ \\ 
4 &    b33 &  17-JUL-1991 & 21.0 &   0.079 & 0.000 & $  66^{+  67}_{ -52}$ \\ 
5 &    b34 &   7-AUG-1991 & 21.0 &   0.064 & 0.043 & $ -17^{+  57}_{ -40}$ \\ 
6 &    b35 &  28-AUG-1991 & 22.3 &   0.035 & 0.024 & $  56^{+  74}_{ -60}$ \\ 
7 &    b36 &  19-SEP-1991 & 19.7 &   0.000 & 0.000 & $  82^{+  59}_{ -45}$ \\ 
8 &    b38 &  10-OCT-1991 & 19.9 &   0.068 & 0.059 & $  73^{+  76}_{ -63}$ \\ 
9 &    b39 &  30-OCT-1991 & 21.0 &   0.058 & 0.003 & $ 133^{+  87}_{ -68}$ \\ 
10 &    b41 &  21-NOV-1991 & 19.9 &   0.218 & 0.114 & $  40^{+  81}_{ -65}$ \\ 
11 &    b42 &  11-DEC-1991 & 28.0 &   0.098 & 0.010 & $  80^{+  71}_{ -58}$ \\ 
12 &    b45 &  29-JAN-1992 & 21.0 &   0.034 & 0.032 & $  19^{+  58}_{ -43}$ \\ 
13 &    b47 &  20-FEB-1992 & 19.8 &   0.028 & 0.020 & $ 106^{+  69}_{ -54}$ \\ 
14 &    b49 &  12-MAR-1992 & 18.8 &   0.092 & 0.000 & $ -12^{+  52}_{ -31}$ \\ 
15 &    b50 &  31-MAR-1992 & 29.0 &   0.008 & 0.018 & $ 115^{+  66}_{ -54}$ \\ 
 \hline \hline
\multicolumn{2}{|c}{\sc Gallex II} &  \multicolumn{2}{|c|}{Exposure}  &  \multicolumn{3}{c|}{pulse shape analysis} \\ 
\multicolumn{2}{|c|}{Runs}         &  start  & duration (d)          &  $b_L$  & $b_K$ & $P_\odot$ (SNU)  \\ \hline 
16 &    a59 &  19-AUG-1992 & 28.0 &   0.046 & 0.018 & $ 120^{+  68}_{ -56}$ \\ 
17 &    a61 &  17-SEP-1992 & 27.0 &   0.034 & 0.019 & $ 138^{+  64}_{ -53}$ \\ 
18 &    a63 &  15-OCT-1992 & 27.0 &   0.059 & 0.016 & $ 146^{+  66}_{ -54}$ \\ 
19 &    a65 &  12-NOV-1992 & 27.0 &   0.038 & 0.000 & $  38^{+  44}_{ -29}$ \\ 
20 &    a67 &  10-DEC-1992 & 27.0 &   0.000 & 0.000 & $ 123^{+  54}_{ -42}$ \\ 
21 &    a69 &   7-JAN-1993 & 27.0 &   0.051 & 0.021 & $  48^{+  46}_{ -35}$ \\ 
22 &    a71 &   4-FEB-1993 & 27.0 &   0.083 & 0.037 & $  77^{+  52}_{ -41}$ \\ 
23 &    a73 &   4-MAR-1993 & 29.0 &   0.016 & 0.012 & $ 114^{+  58}_{ -46}$ \\ 
24 &    a75 &   3-APR-1993 & 25.0 &   0.035 & 0.024 & $ 151^{+  70}_{ -58}$ \\ 
25 &    a77 &  29-APR-1993 & 27.0 &   0.044 & 0.038 & $   3^{+  44}_{ -29}$ \\ 
26 &    a79 &  27-MAY-1993 & 27.0 &   0.036 & 0.026 & $  59^{+  55}_{ -42}$ \\ 
27 &    a81 &  24-JUN-1993 & 27.0 &   0.040 & 0.017 & $  80^{+  54}_{ -42}$ \\ 
28 &    a83 &  22-JUL-1993 & 27.0 &   0.057 & 0.006 & $  43^{+  43}_{ -31}$ \\ 
29 &    a85 &  19-AUG-1993 & 27.0 &   0.014 & 0.006 & $ 101^{+  51}_{ -40}$ \\ 
30 &    a87 &  16-SEP-1993 & 27.0 &   0.029 & 0.042 & $  37^{+  43}_{ -33}$ \\ 
31 &    a89 &  14-OCT-1993 & 27.0 &   0.019 & 0.038 & $  82^{+  51}_{ -40}$ \\ 
32 &    a91 &  11-NOV-1993 & 27.0 &   0.042 & 0.025 & $  11^{+  37}_{ -25}$ \\ 
33 &    a93 &   9-DEC-1993 & 27.0 &   0.014 & 0.021 & $  37^{+  51}_{ -36}$ \\ 
34 &    a95 &   6-JAN-1994 & 27.0 &   0.024 & 0.011 & $ 108^{+  56}_{ -45}$ \\ 
35 &    a97 &   3-FEB-1994 & 27.0 &   0.032 & 0.018 & $  92^{+  54}_{ -42}$ \\ 
36 &    a99 &   3-MAR-1994 & 27.0 &   0.021 & 0.010 & $  41^{+  47}_{ -34}$ \\ 
37 &   a101 &  31-MAR-1994 & 27.0 &   0.034 & 0.014 & $ 102^{+  51}_{ -41}$ \\ 
38 &   a103 &  28-APR-1994 & 27.0 &   0.056 & 0.014 & $  81^{+  44}_{ -35}$ \\ 
39 &   a105 &  26-MAY-1994 & 27.0 &   0.036 & 0.020 & $ 135^{+  70}_{ -56}$ \\ 
 \hline
\end{tabular}
\caption{Single solar run results with stat. error ($1\sigma$) for {\sc Gallex} I and II. \label{results12} }
\end{center}
\end{table}

\begin{table}[h]
\begin{center}
\renewcommand{\arraystretch}{1.3}
 \footnotesize
\begin{tabular}{|cc|cc|ccc|} \hline
\multicolumn{2}{|c}{\sc Gallex III} &  \multicolumn{2}{|c|}{Exposure}  &  \multicolumn{3}{c|}{pulse shape analysis} \\ 
\multicolumn{2}{|c|}{Runs}         &  start  & duration (d)          &  $b_L$  & $b_K$ & $P_\odot$ (SNU)  \\ \hline 
40 &   a119 &  12-OCT-1994 & 21.0 &   0.058 & 0.011 & $ 173^{+  66}_{ -55}$ \\ 
41 &   a120 &   2-NOV-1994 & 21.0 &   0.031 & 0.007 & $  65^{+  46}_{ -34}$ \\ 
42 &   a121 &  23-NOV-1994 & 21.0 &   0.028 & 0.010 & $  56^{+  41}_{ -32}$ \\ 
43 &   a123 &  15-DEC-1994 & 27.0 &   0.039 & 0.036 & $  47^{+  44}_{ -35}$ \\ 
44 &   a124 &  11-JAN-1995 & 28.0 &   0.079 & 0.049 & $ -28^{+  30}_{ -22}$ \\ 
45 &   a125 &   8-FEB-1995 & 28.0 &   0.039 & 0.021 & $  64^{+  51}_{ -41}$ \\ 
46 &   a127 &   9-MAR-1995 & 29.0 &   0.038 & 0.000 & $  53^{+  37}_{ -26}$ \\ 
47 &   a128 &   7-APR-1995 & 26.0 &   0.030 & 0.000 & $  25^{+  32}_{ -20}$ \\ 
48 &   a129 &   3-MAY-1995 & 28.0 &   0.067 & 0.036 & $   7^{+  42}_{ -32}$ \\ 
49 &   a131 &   1-JUN-1995 & 27.0 &   0.042 & 0.016 & $  90^{+  62}_{ -48}$ \\ 
50 &   a132 &  28-JUN-1995 & 28.0 &   0.058 & 0.017 & $  55^{+  51}_{ -38}$ \\ 
51 &   a133 &  26-JUL-1995 & 28.0 &   0.014 & 0.000 & $  29^{+  32}_{ -20}$ \\ 
52 &   a135 &  24-AUG-1995 & 20.0 &   0.010 & 0.020 & $  27^{+  36}_{ -25}$ \\ 
53 &   a136 &  13-SEP-1995 & 21.0 &   0.027 & 0.013 & $  56^{+  44}_{ -34}$ \\ 
 \hline
\multicolumn{7}{c}{ } \\ \hline
\multicolumn{2}{|c}{\sc Gallex IV} &  \multicolumn{2}{|c|}{Exposure}  &  \multicolumn{3}{c|}{pulse shape analysis} \\ 
\multicolumn{2}{|c|}{Runs}         &  start  & duration (d)          &  $b_L$  & $b_K$ & $P_\odot$ (SNU)  \\ \hline 
54 &   a146 &  14-FEB-1996 & 21.0 &   0.135 & 0.015 & $ 104^{+  61}_{ -48}$ \\ 
55 &   a148 &   7-MAR-1996 & 22.0 &   0.010 & 0.053 & $  47^{+  62}_{ -48}$ \\ 
56 &   a149 &  29-MAR-1996 & 19.0 &   0.053 & 0.012 & $  60^{+  55}_{ -40}$ \\ 
57 &   a151 &  18-APR-1996 & 20.0 &   0.019 & 0.033 & $  28^{+  60}_{ -40}$ \\ 
58 &   a157 &  27-JUN-1996 & 20.0 &   0.063 & 0.020 & $  68^{+  65}_{ -50}$ \\ 
59 &   a158 &  17-JUL-1996 & 21.0 &   0.025 & 0.019 & $  91^{+  68}_{ -52}$ \\ 
60 &   a161 &  29-AUG-1996 & 20.0 &   0.105 & 0.061 & $ -98^{+  52}_{ -43}$ \\ 
61 &   a162 &  18-SEP-1996 & 22.0 &   0.041 & 0.000 & $ 100^{+  59}_{ -44}$ \\ 
62 &   a163 &  10-OCT-1996 & 41.0 &   0.062 & 0.012 & $ 125^{+  59}_{ -49}$ \\ 
63 &   a165 &  21-NOV-1996 & 20.0 &   0.024 & 0.009 & $ 106^{+  65}_{ -51}$ \\ 
64 &   a166 &  11-DEC-1996 & 29.0 &   0.053 & 0.000 & $ 201^{+  69}_{ -58}$ \\ 
65 &   a167 &   9-JAN-1997 & 13.0 &   0.025 & 0.015 & $  82^{+  64}_{ -47}$ \\ 
 \hline
\end{tabular}
\caption{Single solar run results with stat. error ($1\sigma$) for {\sc Gallex} III and IV. \label{results34} }
\end{center}
\end{table}
\twocolumn
\noindent periods I, II and III are in good agreement, the difference for period IV is remarkable. The statistical error bars have a small overlap, but one should keep in mind that both results were derived from the same data set and should be strongly correlated. To estimate the correlation in a quantitative way we compared the single run results of the periods I, II and III. The correlation coefficient $r_{x,y}$ is defined as
\begin{equation}
 r_{xy} = \frac{{\rm cov} (x, y)}{\sigma_x \sigma_y} 
\end{equation}
with the covariance
\begin{equation} {\rm cov} (x,y) = \frac{1}{n-1} \sum_{i=1}^{n} (x_i - \bar x)(y_i - \bar y) \, .
\end{equation}
One gets $r_{xy}=0.826$ and therefore $r^2_{x,y}=0.682$, where the latter is conventionally 
interpreted as the part of the
variance of $x$ caused by changes in $y$ (and vice versa). 
If one applies this expectation to the {\sc Gallex} IV results, only a third of the variation 
is caused by statistical fluctuations. From this point of view the difference between the two 
results is very unlikely.

\begin{figure}[t]
\includegraphics[angle=0,width=\figwid]{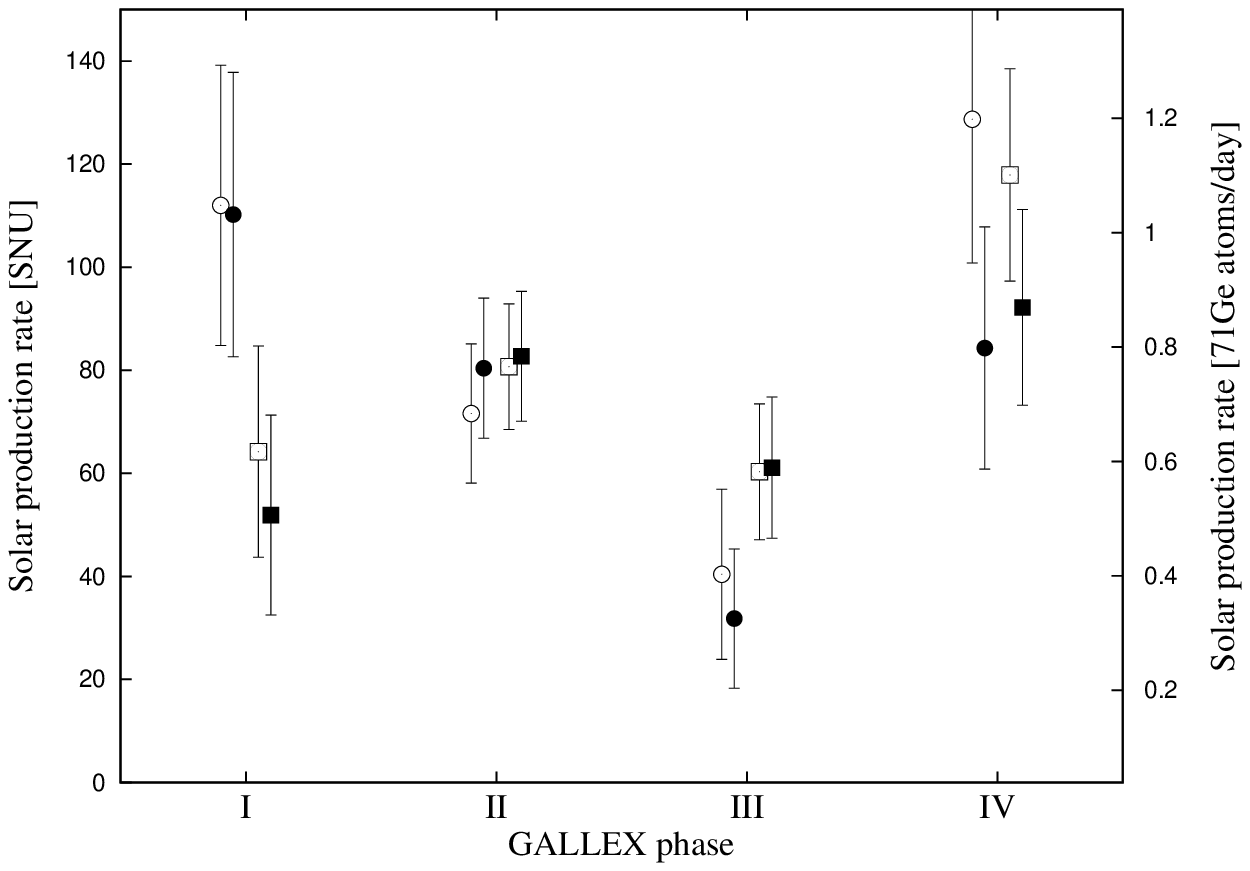}
\caption{Independent analysis for L ($\scriptstyle \odot$, $\bullet$) and K ($\scriptstyle \boxdot$, $\scriptstyle \blacksquare$) events with rise time and pulse shape method, respectively.\label{LK}}
\end{figure}

\begin{figure}[b]
\includegraphics[angle=0,width=\figwid]{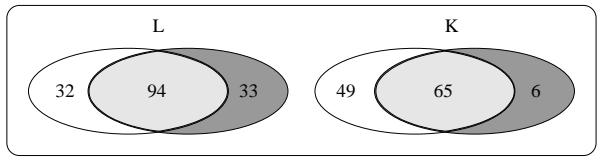}
\caption{Number of selected {\sc Gallex} IV events with rise time (white) and pulse shape (grey) analysis for  L an K energy region.  \label{mengen}}
\end{figure}
Concerning the rise time results it was already noted in \cite{Gx499}\cite{Kir99} that the scattering of the four results is unusual. A $\chi^2$-test for compatibility with a constant mean yields a probability of less than 1\% ($\chi^2=12.7$ with 3 degrees of freedom, assuming symmetric errors). However, it was shown that the scattering is decreased if  different kinds of grouping are applied (e.g. four random divisions) resulting in probabilities up to 26.7\%.
For the results obtained by pulse shape analysis one calculates $\chi^2=7.1$ corresponding to a probability of 7\%, mainly due to the lower {\sc Gallex} IV value. 

As already discussed in \cite{Gx499} eight of the twelve runs of the {\sc Gallex} IV period had problems with electronic noise, which led to a missing baseline in  case of low energy (L) events.
While the uncertainty of the rise time determination increases, the evaluation of pulse shape parameters described in section \ref{psa} is not or only weakly affected by the location of the  baseline level. However, a separate analysis of L and K events reveals that the high {\sc Gallex} IV result obtained with the rise time method cannot exclusively be assigned to the L events (see figure \ref{LK}).

\begin{figure}[t]
\includegraphics[angle=0,width=7.7cm]{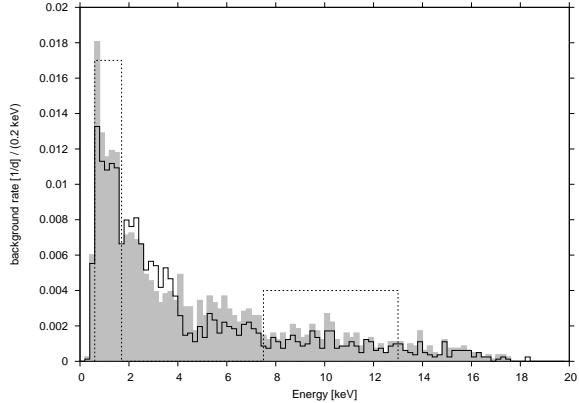}
\caption{Background spectrum after pulse shape cut (black line) and
rise time cut (grey). The energy regions of interest (L and K) are
shown schematically with dashed boxes. \label{bck}}
\end{figure}
The event selection with the pulse shape analysis is more stringent compared to the rise time analysis, therefore it provides a better background reduction (see Figure \ref{bck}) at the cost of a lower cut efficiency especially for K events. 
The diagram in Figure \ref{mengen} shows the number of events selected by both types of analysis for {\sc Gallex} IV.
The difference in the number of accepted K events is as expected, but it is remarkable that the number of accepted L events is almost equal. Therefore, the lower {\sc Gallex} IV 
result is caused by the time distribution of accepted events.

\subsection{Combination with GNO}

After the end of {\sc Gallex} the gallium neutrino observation at
{\sc Lngs} was continued by the GNO collaboration that performed 58
solar runs between 1998 and 2003 \cite{Gno00}\cite{Gno05}. The
experimental setup was basically the same as for {\sc Gallex} except for the electronics,
which had been redesigned in order to replace and modernise the {\sc Gallex}
counting system. The event selection was based on a pulse shape
analysis in which a theoretical pulse shape was fitted to the measured pulse. A
neural network trained by a large amount of reference events
decided on the basis of the fit parameters whether an event was
accepted or rejected \cite{Pan04}.

\begin{figure}[t]
\includegraphics[angle=0,width=\figwid]{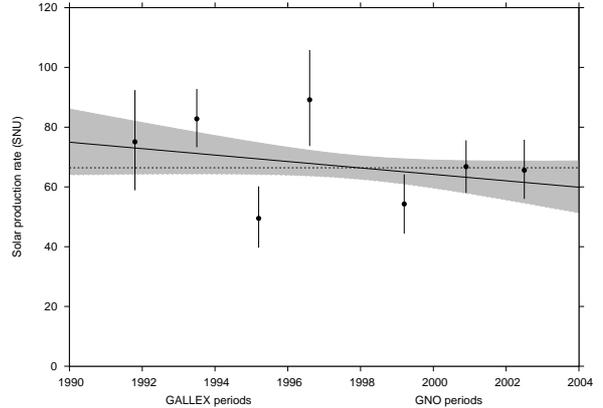}
\caption{Constant and linear fit to the {\sc Gallex}-GNO periods with $1 \sigma$ error region for the linear fit (grey). \label{gnochi}}
\end{figure}

The results of the three GNO measuring periods are shown in
Figure \ref{gallex+gno} together with the four {\sc Gallex} periods.
A $\chi^2$-test for compatibility with a constant mean yields
$\chi^2=9.45$ corresponding to a probability of 15.0\% (6 degrees of
freedom). Since the GNO results seems to have a tendency to smaller
values, we have also performed a linear fit to all seven {\sc
Gallex}-GNO periods, but there was no improvement (the probability even decreased, see Table \ref{chi2-tab} and Figure \ref{gnochi}).

\begin{table}[t]
\center{ \renewcommand{\arraystretch}{1.3}
\footnotesize
\begin{tabular}{c|ccccc}
Fit               & $m$             & $c$     & $\chi^2$ & d.o.f &  $p$ \\ \hline 
$y(t)=c$          &                 & $66.4  $ & 9.44 & 6     & 15.0\%  \\ 
$y(t)=m\bar{t}+c$ & $-1.08 $ & $66.4 $ & 8.55  & 5    & 12.8\%
\end{tabular}
}
\caption{$\chi^2$-fits to the seven {\sc Gallex}-GNO periods for
both a constant and a linear dependence (where $\bar t$ is the average time). \label{chi2-tab} }
\end{table}%

The total GNO result is
\begin{equation}   P_\odot = \Big[ \,  62.9^{+5.5}_{-5.3}
\, {\rm (stat.)} \pm 2.5 \,   {\rm (syst.}) \, \Big] \, {\rm SNU} \;
. \label{gno-result}  \end{equation}
and a combination with the {\sc Gallex} pulse shape result from eq.
\ref{psolar-psa} yields
\begin{equation}   P_\odot = \Big[ \,  67.6 \pm 4.0 \, {\rm (stat.)}
\pm 3.2 \,   {\rm (syst.}) \, \Big] \, {\rm SNU} \; .
\label{gallex-gno-kombi}  \end{equation}
The combination was calculated as a weighted mean using the
statistical errors (with the approximation of symmetry). The
systematic error was obtained by a quadratic combination of both
single errors.

\section{Source experiments}


For a complete test of the experimental performance the {\sc Gallex} collaboration arranged two source experiments \cite{GxS195}\cite{GxS298} in between the solar periods II-III and III-IV respectively. Two intense $^{51}\rm Cr$ neutrino sources were produced by neutron capture on $^{50} \rm Cr$ by irradiation  of isotopically enriched chromium
in the core of the Silo\'{e} reactor in Grenoble. The energies of the emitted neutrinos are about 750 keV (90\%) and 430 keV (10\%). For an accurate knowledge of the source activities $A$ the latter were determined  by different methods (for details see \cite{GxS298}). With the theoretical capture cross section of gallium $\sigma = 58.1^{+2.1}_{-1.6} \times 10^{-46} \, \rm cm^2$ \cite{Bah97} one can predict the expected neutrino signal to compare it with the measurement. The sources were placed in a tube inside the gallium
tank for exposure times of a few days up to 4 weeks. Else, the
experimental procedure was the same as for solar runs.

\begin{table}[t]
\center{
\renewcommand{\arraystretch}{1.3}
\footnotesize
\begin{tabular}{ccc|c|c}
 &     &       &  rise time & pulse shape\\ 
 & Run & date &  P [1/d] &  P [1/d] \\ \hline
1 &   s107 &   23-JUN-1994   &   $11.2^{+ 3.3}_{-2.9}$ &  $12.9^{+ 3.4}_{-3.0}$ \\ 
2 &   s108 &   27-JUN-1994   &   $11.7^{+ 2.9}_{-2.6}$ &  $ 9.9^{+ 2.8}_{-2.4}$ \\ 
3 &   s109 &    1-JUL-1994   &   $ 8.3^{+ 2.4}_{-2.2}$ &  $ 8.1^{+ 2.5}_{-2.2}$ \\ 
4 &   s110 &    6-JUL-1994   &   $ 8.1^{+ 2.0}_{-1.8}$ &  $ 8.2^{+ 2.0}_{-1.8}$ \\ 
5 &   s111 &   13-JUL-1994   &   $ 6.8^{+ 2.0}_{-1.7}$ &  $ 7.5^{+ 2.0}_{-1.8}$ \\ 
6 &   s112 &   20-JUL-1994   &   $ 3.9^{+ 1.6}_{-1.4}$ &  $ 3.8^{+ 1.5}_{-1.3}$\\ 
7 &   s113 &   27-JUL-1994   &   $ 5.1^{+ 1.4}_{-1.3}$ &  $ 4.5^{+ 1.4}_{-1.2}$\\ 
8 &   s114 &    9-AUG-1994   &   $ 2.8^{+ 1.3}_{-1.1}$ &  $ 2.2^{+ 1.2}_{-1.0}$ \\ 
9 &   s115 &   24-AUG-1994   &   $ 3.1^{+ 1.2}_{-1.1}$ &  $ 1.8^{+ 1.1}_{-0.9}$  \\ 
10 &  s116 &    7-SEP-1994   &   $ 0.3^{+ 0.7}_{-0.5}$ & $ 0.3^{+ 0.7}_{-0.5}$\\ 
11 &  s117 &   28-SEP-1994   &   $ 1.8^{+ 1.0}_{-0.8}$ & $ 1.6^{+ 1.0}_{-0.8}$ \\ \hline
\multicolumn{3}{c|}{combined source exp. 1: $P(t=0)$ }& $ 11.7 \pm 1.1 $      & $ 11.2 \pm 1.1$ \\ \hline \hline
1 &   s138 &    5-OCT-1995  &   $ 9.8^{+ 3.0}_{-2.6}$ &  $ 9.7^{+ 3.1}_{-2.6}$\\ 
2 &   s139 &    9-OCT-1995  &   $ 9.2^{+ 2.8}_{-2.5}$ &   $ 9.4^{+ 2.9}_{-2.6}$ \\ 
3 &   s140 &   13-OCT-1995  &   $ 7.0^{+ 1.4}_{-1.2}$ &   $ 6.9^{+ 1.4}_{-1.3}$\\ 
4 &   s141 &    1-NOV-1995  &   $ 5.8^{+ 1.3}_{-1.2}$ &   $ 5.6^{+ 1.4}_{-1.2}$\\ 
5 &   s142 &   22-NOV-1995  &   $ 2.0^{+ 1.1}_{-1.0}$ &   $ 2.2^{+ 1.1}_{-0.9}$ \\ 
6 &   s143 &   20-DEC-1995  &   $ 1.6^{+ 1.0}_{-0.8}$ &   $ 2.0^{+ 0.9}_{-0.8}$\\ 
7 &   s144 &   17-JAN-1996  &   $ 1.5^{+ 0.9}_{-0.8}$ &   $ 1.5^{+ 0.9}_{-0.8}$\\ \hline
\multicolumn{3}{c|}{combined source exp. 2: $P(t=0)$ } & $ 10.4^{+ 1.2}_{-1.1}$ &   $ 10.5\pm 1.2$\\ \hline 
\end{tabular}}
\caption{Results of single runs of both {\sc Gallex} $^{51}$Cr source experiments expressed as production rate $P(t=0)$ at the beginning of each run. Errors are 1$\sigma$ (statistical only). \label{quellruns-tab} }
\end{table}

Compared to the previous published results in \cite{GxS298}, the reanalysis of the source experiments  considers the following changes:

\begin{list}{$\bullet$}{\setlength{\topsep}{1ex} \setlength{\itemsep}{-0.3ex} \setlength{\leftmargin3.0ex} }
\item new counter efficiencies due to more precise calibrations (6 of 18 source runs were affected).
\item the update of the solar production rate by the combined result of {\sc Gallex + GNO}, which has to be treated as additional side reaction in the source experiments.
\item event selection with pulse shape analysis instead of rise time. For an easier comparison the rise time results are given, too.
\end{list}

The analysis procedure is the same as for the solar runs except for the time dependence of the source activity. The $^{51}\rm Cr$ half-life of $27.7 \, \rm d$ has to be considered in the likelihood function. 
The single run results are listed in Table \ref{quellruns-tab}. The time scale refers to the end of bombardment of source production, which is also the zero time for the combined analysis of all runs. The resulting source induced production rates $R$ are given in 
Table \ref{quellruns-tab}, too. The corresponding source activities $A^\nu$ can be obtained by considering the cross section, the geometry of the gallium tank and the source positions \cite{GxS298} by
\begin{equation}
 R_1(t)=0.1856 \, A^\nu_1(t) \, , \quad R_2(t)=0.1866 \, A^\nu_2(t) \,  
\end{equation}
where the unit of $R$ is 1/d if $A$ is given in PBq. They are listed in Table \ref{source-tab} together with the ratio $r$ of $A^\nu$ to the expected source activity $A$. 

\begin{table}[t]
\center{ 
\renewcommand{\arraystretch}{1.5} \footnotesize
\begin{tabular}{@{}c|cc|c|}
            &  $A^{(\nu)}$  [PBq]   &  $r = A^\nu/A$           & $r$ \cite{GxS298}         \\ \hline
source 1    &  $63.4^{+1.1}_{-1.6}$ &                          &                        \\
rise time   &  $63.2^{+6.7}_{-6.5}$ &  $0.997^{+0.11}_{-0.11}$ & $1.01^{+0.12}_{-0.11}$ \\
pulse shape &  $60.4^{+6.6}_{-6.3}$ &  $0.953^{+0.11}_{-0.11}$ &                        \\ \hline
source 2    &  $69.1^{+3.3}_{-2.1}$ &                          &                        \\
rise time   &  $55.8^{+6.8}_{-6.6}$ &  $0.807^{+0.11}_{-0.10}$ & $0.84^{+0.12}_{-0.11}$ \\
pulse shape &  $56.1^{+7.0}_{-6.7}$ &  $0.812^{+0.10}_{-0.11}$ &                        \\ \hline \hline
rise time   &                       &  $0.902 \pm 0.078$       & $0.93 \pm 0.08$        \\
pulse shape &                       &  $0.882 \pm 0.078$       &                        \\ \hline
\end{tabular}}
\caption{Results $A^\nu$ of the source experiments compared to the expected source activity $A$
(both referring to the end of bombardment of the source production). For comparison, the last column gives the results as published before the present reanalysis.  \label{source-tab}}
\end{table}

\subsection{Discussion of the source experiments}

We know from the $^{71}$As experiment performed at the end of {\sc
Gallex} that the Ge extraction yield is very close to 100\%. Since
the ground-state to ground-state cross section is known to within
1\%, this implies that the two $^{51}$Cr source experiments performed
in {\sc Gallex} have measured the contribution of the first two
excited states in $^{71}$Ge to the $^{71}$Ga neutrino capture cross
section. Reanalysing the data from these two source experiments using
the pulse shape discrimination and improved counting efficiencies
yields $r = 0.882 \pm 0.078$ (see Table \ref{source-tab}).
This ratio is $1.5\sigma$ away from the expectation value 1.0 where
a 5\% contribution from the first two excited states is included.

If the results from the $^{51}$Cr and $^{37}$Ar source experiments 
performed in the frame of {\sc Sage} \cite{Sag99}\cite{Sag06} are also included, the total
ratio is $0.87 \pm 0.06$ (though an experiment equivalent to the {\sc
Gallex} $^{71}$As experiment has not been performed for {\sc Sage}).
This low value indicates that the contribution of the
first two excited states is rather small. This is in agreement with
the finding by Hata and Haxton \cite{Hat95} that the assumed
proportionality between (p,n) forward scattering cross sections and
Gamow-Teller strength is not always valid for weaker GT transitions.

If it is adopted that the excited state contribution to the $^{51}$Cr
cross section is closer to 0\%  than to 5\% as estimated by Bahcall, then
this is also true for the $^{7}$Be neutrino capture cross section
where the assumed contribution is 6\% according to Bahcall
\cite{Bah97} (derived from the (p,n) experiments). As a consequence
the $^7$Be contribution of 
$34.8^{+4.8}_{-4.3}$ SNU \cite{Bah04a} 
to the total solar neutrino
capture (without oscillations) should be reduced to 32.7 SNU 
with a slightly reduced error.

\section*{Acknowledgements}
We thank all members of the {\sc Gallex} Collaboration (\cite{Gx499}-\cite{Gx396}) for their respective
contributions towards producing the {\sc Gallex} data.



\end{document}